\documentclass[preprint,prd,nofootinbib,tightenlines,groupedaddress,amsmath,amssymb]{revtex4}

\usepackage{bm}
\usepackage{color}
\usepackage{graphicx}
\allowdisplaybreaks
\usepackage{multirow}

\begin{document}


\title{Predictive $CP$ Violating Relations for Charmless Two-body  Decays of Beauty Baryons $\Xi^{-,\;0}_b$ and $\Lambda_b^0$ With Flavor $SU(3)$ Symmetry}

\author{Xiao-Gang He$^{1,2,3}$\footnote{hexg@phys.ntu.edu.tw}}
\author{Guan-Nan Li${}^{2}$\footnote{lgn198741@126.com}}

\affiliation{${}^{1}$INPAC, SKLPPC and Department of Physics,
Shanghai Jiao Tong University, Shanghai, China}
\affiliation{${}^{2}$CTS, CASTS and
Department of Physics, National Taiwan University, Taipei, Taiwan}
\affiliation{${}^{3}$Physics Division, National Center for
Theoretical Sciences, Hsinchu, Taiwan}

\date{\today $\vphantom{\bigg|_{\bigg|}^|}$}

\date{\today}

\vskip 1cm
\begin{abstract}
Several baryons containing a heavy b-quark have been discovered.  The decays of these states provide new platform for testing the standard model (SM). We study $CP$ violation in SM for charmless two-body decays of the flavor $SU(3)$ anti-triplet beauty baryon (b-baryon) ${\cal B} = (\Xi^-_b,\;\Xi^0_b,\;\Lambda_b^0)$ in a model independent way.
We found, in the flavor $SU(3)$ symmetry limit, a set of new predictive relations among the branching ratio $Br$ and $CP$ asymmetry $A_{CP}$ for $\cal B$ decays,  such as $A_{CP}(\Xi_b^-\to K^0 \Xi^-)/A_{CP}(\Xi_b^-\to \bar K^0 \Sigma^-) = - Br(\Xi_b^- \to \bar K^0 \Sigma^-)/Br(\Xi_b^- \to K^0 \Sigma^-)$,
$A_{CP}(\Lambda_b^0 \to \pi^- p)/A_{CP} (\Xi^0_b \to K^- \Sigma^+) =
-Br(\Xi^0_b \to K^- \Sigma^+)\tau_{\Lambda_b^0}/Br(\Lambda_b^0 \to \pi^- p)\tau_{\Xi^0_b}$,
and
$A_{CP}(\Lambda_b^0 \to K^- p)/ A_{CP} (\Xi^0_b \to \pi^- \Sigma^+) =
-Br(\Xi^0_b \to \pi^- \Sigma^+)\tau_{\Lambda_b^0}/ Br(\Lambda_b^0 \to K^- p)\tau_{\Xi^0_b}$.
Future data from LHCb can test these relations and also other relations found.
\end{abstract}

\pacs{PACS numbers: }

\maketitle


Several baryons containing a heavy b-quark, the beauty baryon (b-baryon) $\cal B$,  have been discovered~\cite{pdg}.  The study of heavy mesons containing a b-quark, the $B$ mesons, provided crucial information~\cite{pdg} in establishing the standard model (SM) for $CP$ violation, the Cabibbo-Kobayashi-Maskawa (CKM) model~\cite{km}. The decays of the $\cal B$ b-baryons  will, with no doubt, provide a new platform to further test the CKM model of CP violation~\cite{cdf1,model,gronau}. New data on $\cal B$ b-baryon will continue come from the LHCb. It is timely to investigate ways to test $CP$ violation in the SM using $\cal B$ b-baryon decays.

For $CP$ violation studies, rare charmless decays of $\cal B$ can play an important role because in these decays both tree and loop level contributions are substantial, providing the possibility of having large $CP$ asymmetries~\cite{cdf1,model}. We will consider such decays. Due to our poor understanding of low energy $QCD$, the evaluations of the decay amplitudes are pluged  with large uncertainties. Flavor $SU(3)$ symmetry has been shown to be an excellent tool in reducing uncertainties by obtain relations among different decays for particles containing a b-quark~\cite{su3}. Several relations obtained for $B$ meson decays have been tested to good precisions, in particular for two-body charmless $B$ meson decays~\cite{he1,he2,he3,he4}. With more particles in the final states, the analysis become more complicated and large flavor $SU(3)$ breaking uncertainties become difficult to control~\cite{he5}.
In this letter we will study $CP$ violating relations  for low-lying ${1\over 2}^+$  $\cal B$ b-baryon states decay into two charmless light particles using flavor $SU(3)$ symmetry.

The low-lying ${1\over 2}^+$ $\cal B$ b-baryons contain a flavor $SU(3)$ anti-triplet and a sextet~\cite{heavy-baryon}. We concentrate on the anti-triplet decays. The anti-triplet $\cal B$ b-baryons will be indicated by
 \begin{eqnarray}
 ({\cal B}_{\bar {3}})_{ij}=\left ( \begin{array}{ccc}0&\Lambda^{0}_{b}&\Xi^{0}_{b}\\
 -\Lambda^{0}_{b}&0&\Xi^{-}_{b}\\
 -\Xi^{0}_{b}&-\Xi^{-}_{b}&0
 \end{array}
\right )
 \end{eqnarray}
Their quark compositions are~\cite{heavy-baryon}
\begin{eqnarray}
  \Lambda^{0}_{b}={1\over\sqrt{2}}(ud-du)b;\; \Xi^{0}_{b}={1\over\sqrt{2}}(us-su)b;\; \Xi^{-}_{b}={1\over\sqrt{2}}(ds-sd)b\;.
\end{eqnarray}

The two charmless states in the final state of $\cal B$ decay are the ${1\over 2}^+$   baryon $P$ in the octet $\cal F$ and the pseudoscalar meson $M$ in the octet $\cal M$, respectively. They are
\begin{eqnarray}
{\cal M} = \left ( \begin{array}{ccc}
{\pi^{0}\over \sqrt{2}}+{\eta_{8} \over \sqrt{6}}& \pi^{+}& K^{+}\\
\pi^{-}& -{\pi^{0}\over \sqrt{2}}+{\eta_{8} \over \sqrt{6}}&{K}^{0}\\
K^{-}&  \bar K^{0} & -{2\eta_{8} \over \sqrt{6}}
\end{array}
\right )\;,\;\;
{\cal F} = \left ( \begin{array}{ccc}
{\Sigma^{0}\over \sqrt{2}}+{\Lambda^{0} \over \sqrt{6}}& \Sigma^{+}& p\\
\Sigma^{-}& -{\Sigma^{0}\over \sqrt{2}}+{\Lambda^{0} \over \sqrt{6}}&n\\
\Xi^{-}&  \Xi^{0} & -{2\Lambda^{0} \over \sqrt{6}}
\end{array}
\right )\;.
\end{eqnarray}

The $\cal B \to \cal M + \cal F$ decay can be induced by weak interaction in the SM and can have both parity conserving $A_c$ and violating $A_v$ amplitudes in the form ${\cal M} \bar {\cal F} (A_v + iA_c \gamma_5) \cal B$. This leads to
a decay width given by
\begin{eqnarray}
\Gamma&=&2\vert p_{c}\vert (\vert{\cal S}\vert^2+\vert {\cal P}\vert^2)\;,
\end{eqnarray}
where $\vert p_{c}\vert =\sqrt{E^2_{\cal F}-m^2_{\cal F}}$. $m_{\cal B}$ and $m_{\cal M}, m_{\cal F}$ are the masses of the initial and final particles.
$E_{\cal F}$ is the energy of the final baryon $\cal F$.
${\cal S}$ and ${\cal P}$ are referred as $S$ and $P$ wave amplitudes with
\begin{eqnarray}
{\cal S}=A_v \sqrt{{(m_{\cal B}+m_{\cal F})^2-m^2_{\cal M} \over  16 \pi m^2_{\cal B}}}\;,\;\;{\cal P}= A_c\sqrt{{(m_{\cal B}-m_{\cal F})^2-m^2_{\cal M} \over 16\pi m^2_{\cal B}}}\;.
\end{eqnarray}

In the SM there are tree and penguin contributions to  ${\cal S}$ and ${\cal P}$ for $\Delta S=0$ and $\Delta S=-1$ processes. The $ {\cal S}$ and ${\cal P}$ amplitudes can be written as:
\begin{eqnarray}
{\cal S}(q)=V_{ub}V^{*}_{uq}T(q)_{0}+V_{tb}V^{*}_{tq}P(q)_{0}\;,\;\;
{\cal P}(q)=V_{ub}V^{*}_{uq}T(q)_{1}+V_{tb}V^{*}_{tq}P(q)_{1}\;,
\end{eqnarray}
where $q$ can be $d$ or $s$. The sub-indices $0,1$  denote the $S$ and $P$ wave amplitudes. $V_{ij}$ is the CKM matrix element.

In the SM, there are relations between the decay amplitudes with $q=d$ and $q=s$ in the flavor $SU(3)$ symmetry limit. A particularly interesting set of relations is the one with $U$-spin symmetry relate CP violation in some $\Delta S = 0$ and $\Delta S=-1$ processes. We now show some details on how to obtain such relations.

In the SM, the effective operator for the decay processes under consideration at one electroweak loop level is given by
\begin{eqnarray}
 H_{eff}^q = {4 G_{F} \over \sqrt{2}} [V_{ub}V^{*}_{uq} (c_1 O_1 +c_2 O_2)
   - \sum_{i=3}^{12}(V_{ub}V^{*}_{uq}c_{i}^{uc} +V_{tb}V_{tq}^*
   c_i^{tc})O_{i}],
\end{eqnarray}
where $q$ can be $d$ or $s$, the coefficients
$c_{1,2}$ and $c_i^{jk}=c_i^j-c_i^k$, with $j$ and $k$ indicate the internal quark,
are the Wilson Coefficients (WC) for the operators composed of quarks, photon and gluon fields.
$O_{1,2}$ , $O_{3, 4,5,6}$ and $O_{7,8,9,10}$ are the tree, penguin and electroweak penguin operators.
$O_{11,12}$ are the photonic and gluonic dipole penguin operators.
Details of the operators and their associated WC have been studied by several groups and can be
found in Ref.~\cite{heff}.
In the above the factor $V_{cb}V_{cq}^*$ has
been eliminated using the unitarity property of the CKM matrix.


At the hadron level, the decay amplitude can be generically written as
\begin{eqnarray}
{\cal A} = \langle {\cal F} {\cal M}\vert H_{eff}^q\vert {\cal {B}}\rangle =  V_{ub}V^*_{uq} T(q) + V_{tb}V^*_{tq}P(q).
\end{eqnarray}

The operators $O_{i}$ contains $ \overline{3}$, $6$ , $\overline{15}$ of flavor $SU(3)$ irreducible representations. Indicating these representations by matrices $ H(\overline{3})$,
$H(6)$, $H(\overline{15})$~\cite{su3,he1}.
The non-zero entries of the matrices $H(i)$ are given as the followed~\cite{su3, he1}.

For $\Delta S=0$,
\begin{eqnarray}
&&H(\overline{3})^{2}=1\;,\; H(6)^{12}_{1}=H(6)^{23}_{3}=1\;,
\;H(6)^{21}_{1}=H(6)^{32}_{3}=-1\;,\nonumber\\
&&H(\overline{15})^{12}_{1}=H(\overline{15})^{21}_{1}=3,\;H(\overline{15})^{22}_{2}=-2\;,
\;H(\overline{15})^{32}_{3}=H(\overline{15})^{23}_{3}=-1\;,
\end{eqnarray}
and for $\Delta S=-1$,
\begin{eqnarray}
&&H(\overline{3})^{3}=1\;,\; H(6)^{13}_{1}=H(6)^{32}_{2}=1\;,
\;H(6)^{31}_{1}=H(6)^{23}_{2}=-1\;,\nonumber\\
&&H(\overline{15})^{13}_{1}=H(\overline{15})^{31}_{1}=3\;,\;H(\overline{15})^{33}_{3}=-2\;,
\;H(\overline{15})^{32}_{2}=H(\overline{15})^{23}_{2}=-1\;.
\end{eqnarray}

For an initial $\cal B$ b-baryon, it is understood that the Hamiltonian will annihilate the b-quark and contract $SU(3)$ indices in an appropriate way with final states $\cal F$ and $\cal M$ to obtain $SU(3)$ invariant amplitudes. As far as $SU(3)$ properties are concerned, the $\cal S$ and $\cal P$ amplitudes will have various $SU(3)$ irreducible contributions which can be obtained from the following invariant amplitudes, taking the tree $\cal S$ amplitude as example

\begin{eqnarray}
T_{tri}(q)&=&a(\overline{3})\langle {\cal F}^{k}_{l}{\cal M}^{l}_{k}\vert H(\overline{3})^{i} \vert {\cal B}_{i'i''}\rangle \epsilon ^{ii'i''}
+b(\overline{3})_{1}\langle {\cal F}^{k}_{j} {\cal M}^{i}_{k} \vert H(\overline{3})^{j}\vert {\cal B}_{i'i''}\rangle \epsilon ^{ii'i''}\nonumber\\
&+&b(\overline{3})_{2}\langle {\cal F}^{i}_{k}{\cal M}^{k}_{j}\vert H(\overline{3})^{j} \vert {\cal B}_{i'i''}\rangle \epsilon ^{ii'i''}
+a(6)_{1}\langle {\cal F}^{k}_{l} {\cal M}^{l}_{j} \vert H(6)^{ij}_{k}\vert {\cal B}_{i'i''}\rangle \epsilon ^{ii'i''}\nonumber\\ &+&a(6)_{2}\langle {\cal F}^{l}_{j}{\cal M}^{k}_{l}\vert H(6)^{ij}_{k}\vert {\cal B}_{i'i''}\rangle \epsilon ^{ii'i''}
+b(6)_{1}\langle {\cal F}^{l}_{k} {\cal M}^{i}_{j}\vert H(6)^{jk}_{l}\vert {\cal B}_{i'i''}\rangle \epsilon ^{ii'i''}\nonumber\\
&+&b(6)_{2}\langle {\cal F}^{i}_{j}{\cal M}^{l}_{k}\vert H(6)^{jk}_{l}\vert {\cal B}_{i'i''}\rangle \epsilon ^{ii'i''}
+a(\overline{15})_{1}\langle {\cal F}^{k}_{l} {\cal M}^{l}_{j}\vert H(\overline{15})^{ij}_{k}\vert {\cal B}_{i'i''}\rangle \epsilon ^{ii'i''}\nonumber\\
&+&a(\overline{15})_{2} \langle {\cal F}^{l}_{j}{\cal M}^{k}_{l}\vert H(\overline{15})^{ij}_{k}\vert {\cal B}_{i'i''}\rangle \epsilon ^{ii'i''}
+b(\overline{15})_{1}\langle  {\cal F}^{l}_{k} {\cal M}^{i}_{j}\vert H(\overline{15})^{jk}_{l}\vert {\cal B}_{i'i''}\rangle \epsilon ^{ii'i''}\nonumber\\
&+&b(\overline{15})_{2}\langle {\cal F}^{i}_{j}{\cal M}^{l}_{k}\vert H(\overline{15})^{jk}_{l}\vert {\cal B}_{i'i''}\rangle \epsilon ^{ii'i''}\nonumber\\
&+&c(\overline{3})\langle {\cal M}^{i}_{j}{\cal F}^{i'}_{j'} \vert H(\overline{3})^{i''}  \vert {\cal B}_{jj'}\rangle \epsilon_{ii'i''}
+d(\overline{3})_{1}\langle {\cal M}^{i}_{j}{\cal F}^{i'}_{j'}\vert H(\overline{3})^{j}\vert {\cal B}_{i''j'}\rangle\epsilon_{ii'i''}\nonumber\\
&+&d(\overline{3})_{2}\langle {\cal F}^{i}_{j}{\cal M}^{i'}_{j'}\vert H(\overline{3})^{j}\vert {\cal B}_{i''j'}\rangle\epsilon_{ii'i''}
+e(\overline{3})_{1} \langle {\cal M}^{i}_{j'}{\cal F}^{i'}_{j}\vert H(\overline{3})^{j}\vert {\cal B}_{i''j'}\rangle\epsilon_{ii'i''}\nonumber\\
&+&e(\overline{3})_{2}\langle {\cal F}^{i}_{j'}{\cal M}^{i'}_{j}\vert H(\overline{3})^{j}\vert {\cal B}_{i''j'} \rangle \epsilon_{ii'i''}
+c(6)\langle {\cal M}^{i}_{j}{\cal F}^{i'}_{j'}\vert H(6)^{jj'}_{k}\vert {\cal B}_{i''k}\rangle\epsilon_{ii'i''}\nonumber\\
&+&d(6)_{1}\langle {\cal M}^{i}_{j}{\cal F}^{i'}_{j'}\vert H(6)^{i''j}_{k}\vert {\cal B}_{j'k}\rangle\epsilon_{ii'i''}
+d(6)_{2} \langle {\cal F}^{i}_{j}{\cal M}^{i'}_{j'}\vert H(6)^{i''j}_{k}\vert {\cal B}_{j'k}\rangle\epsilon_{ii'i''}\nonumber\\
&+&e(6)_{1} \langle {\cal M}^{i}_{j}{\cal F}^{i'}_{j'}\vert H(6)^{i''j'}_{k}\vert {\cal B}_{jk}\rangle\epsilon_{ii'i''}+
e(6)_{2}\langle {\cal F}^{i}_{j}{\cal M}^{i'}_{j'}\vert H(6)^{i''j'}_{k}\vert {\cal B}_{jk}\rangle\epsilon_{ii'i''}\nonumber\\
&+&f(6)\langle {\cal M}^{i}_{j}{\cal F}^{k}_{j'}\vert H(6)^{i'i''}_{k}\vert {\cal B}_{jj'}\rangle\epsilon_{ii'i''}+
g(6) \langle {\cal M}^{k}_{j}{\cal F}^{i}_{j'}\vert H(6)^{i'i''}_{k}\vert {\cal B}_{jj'}\rangle\epsilon_{ii'i''}\nonumber\\
&+&m(6) \langle {\cal M}^{k}_{j}{\cal F}^{j}_{k}\vert H(6)^{ii'}_{l}\vert {\cal B}_{i''l}\rangle\epsilon_{ii'i''}+
n(6)_{1}  \langle {\cal M}^{k}_{j}{\cal F}^{j}_{l}\vert H(6)^{ii'}_{k}\vert {\cal B}_{i''l}\rangle\epsilon_{ii'i''}\nonumber\\
&+&n(6)_{2}\langle  {\cal F}^{k}_{j}{\cal M}^{j}_{l}\vert H(6)^{ii'}_{k}\vert {\cal B}_{i''l}\rangle\epsilon_{ii'i''}
+c(\overline{15})\langle  {\cal M}^{i}_{j}{\cal F}^{i'}_{j'}\vert H(\overline{15})^{jj'}_{k}\vert {\cal B}_{i''k}\rangle\epsilon_{ii'i''}\nonumber\\
&+&d(\overline{15})_{1}\langle  {\cal M}^{i}_{j}{\cal F}^{i'}_{j'}\vert H(\overline{15})^{i''j}_{k}\vert {\cal B}_{j'k}\rangle\epsilon_{ii'i''}
+d(\overline{15})_{2} \langle {\cal F}^{i}_{j}{\cal M}^{i'}_{j'}\vert H(\overline{15})^{i''j}_{k}\vert {\cal B}_{j'k}\rangle\epsilon_{ii'i''}\nonumber\\
&+&e(\overline{15})_{1}\langle {\cal M}^{i}_{j}{\cal F}^{i'}_{j'}\vert H(\overline{15})^{i''j'}_{k}\vert {\cal B}_{jk}\rangle\epsilon_{ii'i''}+
e(\overline{15})_{2}\langle {\cal F}^{i}_{j}{\cal M}^{i'}_{j'}\vert H(\overline{15})^{i''j'}_{k}\vert {\cal B}_{jk}\rangle\epsilon_{ii'i''}\;
\end{eqnarray}

Expanding the above invariant amplitudes, we obtain contributions to individual decay processes.  For example, expressing the tree decay amplitudes in terms of the coefficients in $SU(3)$ invariant amplitudes, we have
\begin{eqnarray}
T(\Lambda^{0}_{b} \to \pi^{-} p)&=&- 2a(6)_{1} -2a(\overline{15})_{1}+2b(\overline{3})_{2} + 2b(6)_2 + 6b(\overline{15})_2
+c(\overline{3})+d(\overline{3})_{1}-e(\overline{3})_{2}-c(6)\nonumber\\&+&d(6)_{2}-e(6)_{1}-2f(6)-2g(6)+2n(6)_{2}+3c(\overline{15})
+2d(\overline{15})_{1}-3d(\overline{15})_{2}+3e(\overline{15})_{1}\nonumber\\&-&2e(\overline{15})_{2}\;;
\\
T(\Lambda^{0}_{b} \to K^{-}p)&=&2a(\overline{3})-2a(6)_{2}-4a(\overline{15})_{1}+6a(\overline{15})_{2}+2b(\overline{3})_{2}+2b(6)_{2}+6b(\overline{15})_{2}
\nonumber\\&+&d(\overline{3})_{1}-e(\overline{3})_{2}-c(6)+d(6)_{1}-e(6)_{2}+2n(6)_{1}+3c(\overline{15})+d(\overline{15})_{1}-e(\overline{15})_{2}
\;;
\nonumber\\
T(\Lambda^{0}_{b} \to \pi^{-} \Sigma^{+})&=&2a(\overline{3})+2a(6)_{1}-2a(6)_{2}-2a(\overline{15})_{1}+6a(\overline{15})_{2}-c(\overline{3})
+b(6)_{1}-b(6)_{2}+c(6)_{1}\nonumber\\&-&c(6)_{2}+2d(6)+2e(6)+2g(6)_{1}-2g(6)_{2}
-d(\overline{15})_{1}+3d(\overline{15})_{2}-3e(\overline{15})_{1}+e(\overline{15})_{2}\;.\nonumber
\end{eqnarray}

We have\cite{gronau}
\begin{eqnarray}
T(\Lambda^{0}_{b} \to K^{-}p)-T(\Lambda^{0}_{b} \to \pi^{-} \Sigma^{+})=T(\Lambda^{0}_{b} \to \pi^{-} p)\;.\label{sum}
\end{eqnarray}

We find several relations among the decay amplitudes shown below
\begin{eqnarray}
&& T(\Xi^{-}_{b} \to K^{-} n)= T(\Xi^{-}_{b} \to \pi^{-} \Xi^{0})\;, \;\;\;\;\;\;\;\;
 T(\Xi^{0}_{b} \to \bar{K}^{0} n)=-T(\Lambda^{0}_{b} \to K^{0} \Xi^{0})\;, \nonumber\\
&& T(\Xi^{-}_{b} \to K^{0} \Xi^{-})=T(\Xi^{-}_{b} \to \bar{K}^{0} \Sigma^{-})\;, \;\;\;\;\;
 T(\Xi^{0}_{b} \to K^{0} \Xi^{0})=-T(\Lambda^{0}_{b} \to \bar{K}^{0} n)\;, \nonumber\\
&& T(\Xi^{0}_{b} \to \pi^{-} \Sigma^{+})=-T(\Lambda^{0}_{b} \to K^{-} p)\;, \;\;\;\;\;\;
 T(\Lambda^{0}_{b} \to \pi^{-} p)=-T(\Xi^{0}_{b}\to K^{-} \Sigma^{+})\;,\nonumber\\
&& T(\Xi^{0}_{b} \to \pi^{+} \Sigma^{-})=-T(\Lambda^{0}_{b} \to K^{+} \Xi^{-})\;,\;\;\;
  T(\Lambda^{0}_{b} \to K^{+} \Sigma^{-})=-T(\Xi^{0}_{b} \to \pi^{+} \Xi^{-})\;,\nonumber\\
&& T(\Xi^{0}_{b} \to K^{-} p)=-T(\Lambda^{0}_{b} \to \pi^{-} \Sigma^{+})\;,\;\;\;\;\;\;
  T(\Xi^{0}_{b} \to K^{+} \Xi^{-})=-T(\Lambda^0_{b} \to \pi^{+} \Sigma^{-})\;.
\end{eqnarray}
The full results are listed in Tables \ref{tab:B1S0} to \ref{tab:B3S1}. The expressions for penguin and also $\cal P$ wave amplitudes are similar. Due to mixing between $\eta_8$ and $\eta_1$, the decay modes with $\eta_8$ in the final sates is not as clean as those with $\pi$ and $K$ in the final state to study. We do not list decay amplitudes with $\eta_8$ in the final states for completeness.

It is interesting to note that the pair of decays related by $U$-spin
$\Lambda^{0}_{b} \to \pi^{-} p$ and $\Xi^{0}_{b} \to K ^{-} \Sigma^+$, and,
$\Lambda^{0}_{b} \to K^{-} p$ and $\Xi^{0}_{b} \to \pi ^{-} \Sigma^+$, respectively, have the same tree and penguin amplitudes, that is $T(d)_{j}=T(s)_{j}$, and $P(d)_{j}=P(s)_{j}$.  For these decays, although the absolute values of the decay widths are different, the rate difference $\Delta(i)=\Gamma(i)-\bar {\Gamma}(\bar i)$ are simply related by
\begin{eqnarray}
\Delta(d)=-\Delta(s)
\end{eqnarray}
In obtaining the above relation, we have used the identity:
$Im(V_{ub}V^{*}_{ud}V^{*}_{tb}V_{td})=-Im(V_{ub}V^{*}_{us}V^{*}_{tb}V_{ts})$.


We list those U-spin related decay rate differences pair  with $\Delta S=0$ and $\Delta S=-1$ in the following
\begin{eqnarray}
&&(1)\; \Delta(\Xi^{-}_{b} \to K^{-} n)= -\Delta(\Xi^{-}_{b} \to \pi^{-} \Xi^{0})\;, \;\;\;\;(2)\; \Delta(\Xi^{0}_{b} \to \bar{K}^{0} n)=-\Delta(\Lambda^{0}_{b} \to K^{0} \Xi^{0})\;, \nonumber\\
&&(3)\; \Delta(\Xi^{-}_{b} \to K^{0} \Xi^{-})=-\Delta(\Xi^{-}_{b} \to \bar{K}^{0} \Sigma^{-})\;,\;(4)\; \Delta(\Xi^{0}_{b} \to K^{0} \Xi^{0})=-\Delta(\Lambda^{0}_{b} \to \bar{K}^{0} n)\;, \nonumber\\
&&(5)\; \Delta(\Xi^{0}_{b} \to \pi^{-} \Sigma^{+})=-\Delta(\Lambda^{0}_{b} \to K^{-} p)\;,\;\;\;\;(6)\;\; \Delta(\Lambda^{0}_{b} \to \pi^{-} p)=-\Delta(\Xi^{0}_{b}\to K^{-} \Sigma^{+})\;,\\
&&(7)\; \Delta(\Xi^{0}_{b} \to \pi^{+} \Sigma^{-})=-\Delta(\Lambda^{0}_{b} \to K^{+} \Xi^{-})\;,\;\;
(8)\; \Delta(\Lambda^{0}_{b} \to K^{+} \Sigma^{-})=-\Delta(\Xi^{0}_{b} \to \pi^{+} \Xi^{-})\,\nonumber\\
&&(9)\; \Delta(\Xi^{0}_{b} \to K^{-} p)=-\Delta(\Lambda^{0}_{b} \to \pi^{-} \Sigma^{+})\;,\;\;\;\;
(10)\; \Delta(\Xi^{0}_{b} \to K^{+} \Xi^{-})=-\Delta(\Lambda^0_{b} \to \pi^{+} \Sigma^{-})\;.\nonumber
\end{eqnarray}

The above relations imply relations for $CP$ asymmetries
\begin{eqnarray}
{A_{CP}({\cal B}_a \to {\cal M} {\cal F} )_{\Delta S = 0}\over A_{CP}({\cal B}_b \to  {\cal M} {\cal F})_{\Delta S = -1}}
= - {Br({\cal B}_b \to {\cal M} {\cal F} )_{\Delta S = -1}\over Br({\cal B}_a \to  {\cal M} {\cal F})_{\Delta S = 0}}\cdot
{{\tau_{{\cal B}_a} } \over  {\tau_{{\cal B}_b }} }\;,
\end{eqnarray}
where $\tau_{a,b}$ indicate the lifetimes of b-baryons ${\cal B}_{a,b}$, $Br$ indicates branching ratio, and $A_{CP}$ indicates the $CP$ asymmetry defined as
\begin{eqnarray}
A_{CP}({\cal B} \to {\cal M}  {\cal F})= {\Gamma({\cal B} \to {\cal M}  {\cal F}) - \Gamma({\bar {\cal B}} \to {\bar {\cal M}} {\bar {\cal F}})
\over \Gamma({\cal B} \to {\cal M}  {\cal F}) + \Gamma({\bar {\cal B}} \to {\bar {\cal M}} {\bar {\cal F}})}\;.
\end{eqnarray}

There are similar relations in $B$ decays into two pseudoscalar octet mesons in flavor $SU(3)$ limit. We take the following two relations for discussion for the reason that there are data available for the relevant decays,
\begin{eqnarray}
{A_{CP}(\bar B^0_s \to K^+ \pi^-)\over A_{CP}(\bar B^0\to K^- \pi^+)}
= - {{Br(\bar B^0 \to K^- \pi^+) \tau_{\bar B^0_s} }\over {Br(\bar B^0_s \to K^+ \pi^-)\tau_{\bar B^0} }}\;,\;\;
{A_{CP}(\bar B^0 \to \pi^+ \pi^-)\over A_{CP}(\bar B^0_s\to K^+ K^-)}
= - {{Br(\bar B^0_s \to K^+ K^-) \tau_{\bar B^0}}  \over   { Br(\bar B^0 \to \pi^+ \pi^-)  \tau_{\bar B^0_s}  }}\;.\nonumber\\
\end{eqnarray}
The present data~\cite{pdg,cdf1,lhcb1} give: $-3.41\pm0.55$ and $3.56\pm0.40$ for the left and right hand sides of the first equation above. These two values agree with the prediction very well. For the second equation, the left hand side is $-2.21\pm 1.78$ and the right hand side is
$5.06\pm 0.59$.  The central values do not agree with the prediction, but agree within allowed error bars at 1$\sigma$ level.

Corresponding to the above relations, for each of them there are two pairs. For the first one,
the two pairs are:
\begin{eqnarray}
{A_{CP}(\Lambda_b^0 \to \pi^- p)\over A_{CP} (\Xi^0_b \to K^- \Sigma^+) }=
-{Br(\Xi^0_b \to K^- \Sigma^+)\tau_{\Lambda_b^0}\over Br(\Lambda_b^0 \to \pi^- p)\tau_{\Xi^0_b}}\;,\;\;{A_{CP}(\Lambda_b^0 \to K^+ \Sigma^-)\over A_{CP} (\Xi^0_b \to \pi^+ \Xi^-) }=
-{Br(\Xi^0_b \to \pi^+ \Xi^-)\tau_{\Lambda_b^0}\over Br(\Lambda_b^0 \to K^+ \Sigma^-)\tau_{\Xi^0_b}}\;.\nonumber\\
\end{eqnarray}

For the second one, we have the two pairs as:
\begin{eqnarray}
{A_{CP}(\Lambda_b^0 \to K^- p)\over A_{CP} (\Xi^0_b \to \pi^- \Sigma^+) }=
-{Br(\Xi^0_b \to \pi^- \Sigma^+)\tau_{\Lambda_b^0}\over Br(\Lambda_b^0 \to K^- p)\tau_{\Xi^0_b}}\;,\;\;{A_{CP}(\Lambda_b^0 \to K^+ \Xi^-)\over A_{CP} (\Xi^0_b \to \pi^+ \Sigma^-) }=
-{Br(\Xi^0_b \to \pi^+ \Sigma^-)\tau_{\Lambda_b^0}\over Br(\Lambda_b^0 \to K^+ \Xi^-)\tau_{\Xi^0_b}}\;.\nonumber\\
\label{relation-b}
\end{eqnarray}

We expect that the similar relations will hold at the same level as their $B\to \cal M\cal M$ counter parts.
One expects that $SU(3)$ symmetry holds are 20 to 30 percent level as seen in Kaon and Hyperon decays.
 To estimate the level of $SU(3)$ breaking effects, we define $r_c =-( A_{CP}(\bar B^0_s \to K^+ \pi^-)/ A_{CP}(\bar B^0\to K^- \pi^+))/
(Br(\bar B^0 \to K^- \pi^+) \tau_{\bar B^0_s} / Br(\bar B^0_s \to K^+ \pi^-)\tau_{\bar B^0} )$ as the measure. In the $SU(3)$ limit, $r_c = 1$. Using experimental data, we have $r_c = 0.96\pm 0.19$. The central value is about 5\% away from 1. The 1$\sigma$ level error bar is about 20\%. This is an indication that $SU(3)$ may work better in systems with a b quark than that for Kaon and Hyperon systems. Whether this is an accidental or $SU(3)$ works better for B decays needs to be understood. The relations found for b-baryons above can provide important clues.

At present, only $\Lambda_b \to \pi^+ p$ and $\Lambda_b^0 \to K^- p$ charmless two body decays have been measured exerimentally\cite{pdg,cdf1,lhcb2}. These data points cannot complete relations predicted in eq.(\ref{relation-b}). Only when charmless two body $\Xi_b^0$ decays are also measured, the predictions can be tested. We urge our experimental colleagues to carry out related measurements to test the SM further.

We would like to point out a particularly interesting relation that
\begin{eqnarray}
{A_{CP}(\Xi_b^- \to K^0 \Xi^-)\over A_{CP} (\Xi^-_b \to \bar K^0 \Sigma^-) }=
-{Br(\Xi^-_b \to \bar K^0 \Sigma^-)\over Br(\Xi_b^- \to K^0 \Xi^-)}\;.
\end{eqnarray}
This relation does not involve the lifetimes of the decaying particle. This fact makes it a potentially good test with less error sources.

 Before concluding, we would like to make a comment on the approximate relation existed between $\bar B^0 \to \pi^- \pi^+$ and $\bar B^0\to  K^- \pi^+$ when annihilation contributions are neglected and the possible corresponding one relating $\Lambda_b^0  \to  \pi^- p$ and $\Lambda_b^0\to  K^- p$.

We refer the contributions proportional to $a(i)_\alpha$ as annihilation contributions in view of the fact that the flavor indices of the initial states are contracted by the indices in the Hamiltonian as if the flavor structure of the initial states are annihilated by the Hamiltonian. Since the initial flavor structures are annihilated by the Hamiltonian, no
flavor information flow directly to the final states implying that the flavor structure of the final states have to be created
completely by the weak interaction, the probability is smaller than those other terms where the initial state flows flavor information directly to the final states. Model calculations agree with this picture~\cite{model}. Similar situation happens for $B \to \cal M\cal M$. There have been studied extensively. Theoretical calculations also agree with the assumption of smallness of annihilation contributions~\cite{he4}. More over experimental data support the assumption that the annihilation contributions are small~\cite{pdg,lhcb1}. Under the small annihilation contribution assumption, one has~\cite{he1,he2}
\begin{eqnarray}
{A_{CP}(\bar B^0 \to \pi^- \pi^+)\over A_{CP}(\bar B^0\to K^- \pi^+)}
\approx - {Br(\bar B^0 \to K^- \pi^+)\over Br(\bar B^0 \to \pi^- \pi^+)}\;,\;\;
{A_{CP}(\bar B^0_s \to K^+ \pi^-)\over A_{CP}(\bar B^0_s\to K^+ K^-)}
\approx - {Br(\bar B^0_s \to K^+ K^-)\over Br(\bar B^0_s \to K^+ \pi^-)}\;.\nonumber\\
\end{eqnarray}
For the first equation above, using PDG data~\cite{pdg}, we find that the left side is given by $-3.78\pm 0.67$ and the right hand side given by $3.82\pm 0.17$. For the second equation, the left side is given by $-2.0\pm 1.6$ and the right hand side is given by $4.71\pm 0.60$. The predicted relations are in agreement with data within error bars. In particular the first equation above gives additional confidence on our assumption.

Naivly, one might identify the corresponding decays of
$\bar B^0 \to \pi^- \pi^+,\; K^- \pi^+$ with $\Lambda_b^0  \to \pi^- p, \;K^- p$, respectively. One therefore might  expect that
$A_{CP}(\Lambda_b^0 \to K^- p)/ A_{CP}(\Lambda_b^0 \to \pi^- p)$ to be approximately equal to
$- Br(\Lambda_b^0 \to \pi^- p)/Br(\Lambda_b^0 \to K^- p)$ when annihilation contributions are neglected. This is, however, not true. One can easily see this by inspecting the relation in eq.(\ref{sum}) and $\Lambda_b \to \pi^- \Sigma^+ $ is not purely annihilation contribution induced decay, unlike $\bar B_s^0 \to \pi^-\pi^+$ for in the case of $B \to \cal{M}\cal{M}$ decays. The difference can be traced back to the fact that although both $(B^-,\; \bar B^0,\; \bar B^0_s)$ and $(\Xi_b^-,\;\Xi_b^0,\;\Lambda_b^0)$ are $SU(3)$ anti-triplet, the b-baryon has two light quarks and there are more ways to pass the initial light quarks to the final states allowing non-annihilation contributions to induce\footnote{We thank M. Gronau and J. Rosner for pointing out this crucial fact to us which we had over looked in the previous version.} $\Lambda_b^0 \to \pi^- \Sigma^+$, but not for $\bar B_s^0 \to \pi^-\pi^+$. Therefore even annihilation contributions are neglected, $A_{CP}(\Lambda_b^0 \to K^- p)/ A_{CP}(\Lambda_b^0 \to \pi^- p)$ is not expected to be approximately equal to
$- Br(\Lambda_b^0 \to \pi^- p)/Br(\Lambda_b^0 \to K^- p)$.

In summary we have studied $CP$ violating relations for flavor $SU(3)$ anti-triplet $\cal B$ b-baryons decay into two charmless light particles. These relations can provide tests for SM with flavor $SU(3)$ and the mechanism for heavy b-baryons decays.
We eagerly wait more precise experimental data from LHCb to  further test these relations.  Similar analysis can be carried out for sixtet b-baryon to charmless two-body decays. Detailed analysis on this will be presented elsewhere.

\begin{acknowledgments}

The work was supported in part by MOE Academic Excellent Program (Grant No: 102R891505) and MOST of ROC, and in part by NNSF(Grant No:11175115) and Shanghai Science and Technology Commission (Grant No: 11DZ2260700) of PRC.

\end{acknowledgments}

\begin{table}[h!]
\begin{center}
\caption{$SU(3)$ decay amplitudes for $\Delta S=0, \Xi^{-}_{b} \to MP$ processes.}
\begin{tabular}{l|ccccccccccc}
\hline\hline
\multirow{3}{*}
 {Decay mode} & $a(\overline{3})$ & $a(6)_{1}$ &  $a(6)_{2}$ & $a(\overline{15})_{1}$ & $a(\overline{15})_{2}$ & $b(\overline{3})_{1}$
 & $b(\overline{3})_{2}$  & $b(6)_{1}$ & $b(6)_{2}$ & $b(\overline{15})_{1}$ & $b(\overline{15})_{2}$\\&$c(\overline{3})$ & $d(\overline{3})_{1}$ & $d(\overline{3})_{2}$ & $e(\overline{3})_{1}$ & $e(\overline{3})_{2}$ &
 $c(6)$ & $d(6)_{1}$ & $d(6)_{2}$ & $e(6)_{1}$ & $e(6)_{2}$& \\ &
 $f(6)$ & $g(6)$ & $m(6)$ & $n(6)_{1}$ & $n(6)_{2}$ & $c(\overline{15})$ & $d(\overline{15})_{1}$ & $d(\overline{15})_{2}$ &$e(\overline{15})_{1}$ & $e(\overline{15})_{2}$&\\
\hline
\multirow{3}{*}
{$\Xi^{-}_{b} \to K^{-} n$} & 2(0 & 0 & 1 & 0 & 3  & 1 & 0 & -1 & 0 & -1 & 0)\\
&1 &0 &1 &-1 &0 &1 &-1 &0 &0 &1 & \\
& -2&-2 &0 &-2 &0 &1 &1 &2 &-2 &-1 & \\
\hline
\multirow{3}{*}
{$\Xi^{-}_{b} \to K^{0} \Xi^{-} $ }&2 (0 & 1& 0 & 3 & 0 & 0 & 1 & 0 & -1 & 0 & -1)\\
&1 &1 &0 &0 &-1 &1 &0 &-1 &1 &0 & \\
&2 &2 &0 &0 &-2 &-1 &2 &1 &-1 &-2 & \\
\hline
\multirow{3}{*}
{$\Xi^{-}_{b} \to  \eta_{8} \Sigma^{-}$ }& ${2 \over \sqrt{6}}$ (0  & 1 & 1 & 3 & 3 & 1 & 1 & 1 & -3 & 3 & 3)\\
&${1 \over \sqrt{6}}$ ( 2& 1& 1& -1& -1&2 &-1 &-1 &1 &1 & \\
&4 &4 &0 &-2 &-2 &0 &3 &3 & -3&-3) & \\
\hline
\multirow{3}{*}
{$\Xi^{-}_{b} \to  \pi^{-} \Lambda^{0}$ }& ${2 \over \sqrt{6}}$ (0 & 1 & 1& 3& 3 & 1 & 1 & -3 & 1 & 3 & 3)\\
&${1 \over \sqrt{6}}$ (2 &1 &1 &-1 &-1 &2 &-1 &-1 &1 &1 & \\
&-4 &-4 &0 &-2 &-2 &0 &3 &3 &-3 &-3) & \\
\hline
\multirow{3}{*}
{$\Xi^{-}_{b} \to  \pi^{-} \Sigma^{0}$ }& ${2 \over \sqrt{2}}$  (0 & 1 & -1 & 3 & -3 & -1 & 1 & -1 & 1 & 5  &3)\\
&${1 \over \sqrt{2}}$  ( 0& 1& -1&1 &-1 &0 &1 &-1 &1 &-1 & \\
&0 &0 &0 &2 &-2 &-2 &1 &-1 &1 &-1 )& \\
\hline
\multirow{3}{*}
{$\Xi^{-}_{b} \to  \pi^{0} \Sigma^{-}$ }& ${2 \over \sqrt{2}}$  (0 & -1 & 1 & -3 & 3 & 1 & -1 & 1 & -1 & 3 & 5)\\
& ${1 \over \sqrt{2}}$  (0& -1& 1& -1& 1&0 &-1 &1 &-1 & 1& \\
&0 &0 &0 &-2 &2 &2 &-1 &1 &-1 &1 & \\
\hline
\hline
\end{tabular}
\label{tab:B1S0}
\end{center}
\end{table}
\begin{table}[h!]
\begin{center}
\caption{$SU(3)$ decay amplitudes for $\Delta S=-1, \Xi^{-}_{b} \to MP$ processes.}
\begin{tabular}{l|ccccccccccc}
\hline\hline
\multirow{3}{*}
 {Decay mode} & $a(\overline{3})$ & $a(6)_{1}$ &  $a(6)_{2}$ & $a(\overline{15})_{1}$ & $a(\overline{15})_{2}$ & $b(\overline{3})_{1}$
 & $b(\overline{3})_{2}$  & $b(6)_{1}$ & $b(6)_{2}$ & $b(\overline{15})_{1}$ & $b(\overline{15})_{2}$\\&$c(\overline{3})$ & $d(\overline{3})_{1}$ & $d(\overline{3})_{2}$ & $e(\overline{3})_{1}$ & $e(\overline{3})_{2}$ &
 $c(6)$ & $d(6)_{1}$ & $d(6)_{2}$ & $e(6)_{1}$ & $e(6)_{2}$& \\ &
 $f(6)$ & $g(6)$ & $m(6)$ & $n(6)_{1}$ & $n(6)_{2}$ & $c(\overline{15})$ & $d(\overline{15})_{1}$ & $d(\overline{15})_{2}$ &$e(\overline{15})_{1}$ & $e(\overline{15})_{2}$&\\
\hline
\multirow{3}{*}
{$\Xi^{-}_{b} \to \pi^{-}  \Xi^{0}$} & 2(0 & 0 & 1 & 0 & 3  & 1 & 0 & -1 & 0 & -1 & 0)\\
&(-1 & 1 & -2 & 2 & -1 & 1 & 2 & -1 & 1 & -2& \\
&2 & 2 & 0 & 4 & -2 & -3 & 0 & -3 & 3 & 0) &\\ \hline
\multirow{3}{*}
{$\Xi^{-}_{b} \to \bar{K}^{0} \Sigma^{-} $ }& 2(0 & 1 & 0 & 3 & 0 & 0 & 1 & 0 & -1 & 0 & -1)\\
&(1 & 1&0 & 0 & -1 & 1 & 0 &-1 & 1 & 0 & \\
&-2 & -2 & 0 & 0 & -2 & -1 & 2 & 1 & -1 & -2) & \\ \hline
\multirow{3}{*}
{$\Xi^{-}_{b} \to  \eta_{8} \Xi^{-}$} & ${2 \over \sqrt{6}}$  (0  & -2 & 1 & -6 & 3 & 1 & -2 & 1 & 0 & 3 & 6)\\
&${1 \over \sqrt{6}}$ (1 &-2 &1 &-1 &2 &-1 &-1 &2 &-2 &1 & \\
&-2 &-2 &0 &-2 &4 &3 &-3 &0 &0 &3) & \\ \hline
\multirow{3}{*}
{$\Xi^{-}_{b} \to  K^{-} \Lambda^{0}$} & ${2 \over \sqrt{6}}$ (0 & 1 & -2& 3& -6 & -2 & 1 & 0 & 1 & 6 & 3)\\
&${1 \over \sqrt{6}}$ (-1 &1 &-2 &2 &-1 &1 &2 & -1&1 &-2 & \\
&2 &2 &0 &4 &-2 &-3 &0 &-3 &3 &0) & \\ \hline
\multirow{3}{*}
{$\Xi^{-}_{b} \to  K^{-} \Sigma^{0}$} & ${2 \over \sqrt{2}} $   (0 & 1 & 0 & 3 & 0 & 0 & 1 & -2 & 1 & 4  &3)\\
&${1 \over \sqrt{2}} $   ( 1&1 &0 &0 &-1 &1 &0 &-1 &1 &0 & \\
& -2& -2&0 &0 &-2 &-1 &2 &1 &-1 &-2) & \\ \hline
\multirow{3}{*}
{$\Xi^{-}_{b} \to  \pi^{0} \Xi^{-}$} & ${2 \over \sqrt{2}}$  (0 & 0 & 1 & 0 & 3 & 1 & 0 & 1 & -2 & 3 & 4)\\
&${1 \over \sqrt{2}}$  (1& 0&1 &-1 &0 &1 &-1 &0 &0 &1 &  \\
& 2& 2& 0&-2 &0 &1 &1 &2 &-2 &-1) & \\
\hline
\hline
\end{tabular}
\label{tab:B1S1}
\end{center}
\end{table}
\begin{table}[h!]
\begin{center}
\caption{$SU(3)$ decay amplitudes for $\Delta S=0, \Xi^{0}_{b} \to MP$ processes.}
\begin{tabular}{l|ccccccccccc}
\hline\hline
\multirow{3}{*}
 {Decay mode} & $a(\overline{3})$ & $a(6)_{1}$ &  $a(6)_{2}$ & $a(\overline{15})_{1}$ & $a(\overline{15})_{2}$ & $b(\overline{3})_{1}$
 & $b(\overline{3})_{2}$  & $b(6)_{1}$ & $b(6)_{2}$ & $b(\overline{15})_{1}$ & $b(\overline{15})_{2}$\\&$c(\overline{3})$ & $d(\overline{3})_{1}$ & $d(\overline{3})_{2}$ & $e(\overline{3})_{1}$ & $e(\overline{3})_{2}$ &
 $c(6)$ & $d(6)_{1}$ & $d(6)_{2}$ & $e(6)_{1}$ & $e(6)_{2}$& \\ &
 $f(6)$ & $g(6)$ & $m(6)$ & $n(6)_{1}$ & $n(6)_{2}$ & $c(\overline{15})$ & $d(\overline{15})_{1}$ & $d(\overline{15})_{2}$ &$e(\overline{15})_{1}$ & $e(\overline{15})_{2}$&\\
\hline
\multirow{3}{*}
{$\Xi^{0}_{b} \to K^{+} \Xi^{-}$} &-2 (1 & -1 & 1 & 3 & -1  & 0 & 0 & 0 & 0 & 0 & 0)\\
&(1 &0 &0 &0 &0 &0 &1 & -1& 1& -1& \\
&2 &2 &0 &2 &-2 &0 &-3 &1 &-1 &3) & \\
\hline
\multirow{3}{*}
{$\Xi^{0}_{b} \to \pi^{+} \Sigma^{-} $ }& -2 (1 & -1& 0 & 3 & -2 & 1 & 0 & 1 & 0 & 3 & 0)\\
&(0 &0 &0 & 0& 0& 1& 0& -1&1 &0 & \\
&0 &0 &0 &0 &-2 &3 &0 &-1 &1 &0) & \\
\hline
\multirow{3}{*}
{$\Xi^{0}_{b} \to K^{-} p$}    & -2( 1&  1  &  -1 &-1  &3 &0 &0 &0 &0 &0 &  0) \\
&(1 &0 &0 &0 &0 &0 &-1 &1 &-1 &1 & \\
&-2 &-2 &0 &-2 &2 &0 &1 &-3 &3 &-1) & \\
\hline
\multirow{3}{*}
{$\Xi^{0}_{b} \to \pi^{-} \Sigma^{+}$}    & -2( 1&  0  &  -1 &-2 &3 &0 &1 &0 &1 &0 &  3) \\
&( 0& -1& 0&0 &1 &1 &-1 &0 &0 &1 & \\
& 0& 0& 0& -2& 0&-3 &-1 &0 &0 &1) & \\
\hline
\multirow{3}{*}
{$\Xi^{0}_{b} \to K^{0} \Xi^{0}$ }   & -2( 1&  0  &  1 &-2 &-1 &0 &1 &0 &-1 &0 &  -1) \\
& (0&-1 &0 &1 &0 &-1 &1 &0 &0 &-1 & \\
&0 &0 &0 &2 &0 &1 &3 &0 &0 &-3 )& \\
\hline
\multirow{3}{*}
{$\Xi^{0}_{b} \to \bar{K}^{0} n$   } & -2( 1&  1  &  0 &-1 &-2 &1 &0 &-1 &0 &-1 &  0) \\
&( 0& 0& -1& 1&0 & -1&0 &1 &-1 &0 & \\
& 0& 0& 0& 0& 2& -1& 0&3 &-3 &0) & \\
\hline
\multirow{3}{*}
{$\Xi^{0}_{b} \to  \eta_{8} \Lambda^{0}$ }& ${-1 \over 3}$ (6  & 3 & 3 & -3 & -3 & 1 & 1 & -3 & -3 & 3 & 3)\\
&${1 \over 6}$( 4& -1& -1& 1& 1&6 &3 &3 &-3 &-3 & \\
&0 &0 &0 &6 &6 &0 &3 &3 &-3 &-3) & \\
\hline
\multirow{3}{*}
{$\Xi^{0}_{b} \to  \eta_{8} \Sigma^{0}$} &${-1 \over \sqrt{3} }$ (0  & -1 & -1 & 5 & 5 & -1 & -1 & -1 & 3 & 5 & -3)\\
&${1 \over 2\sqrt{3} }$ (2 &1 &1 &-1 & -1&2 &-1 &-1 &1 &1 & \\
&4 &4 & 0& -2& -2&8 &3 &-13 &13 &-3) & \\
\hline
\multirow{3}{*}
{$\Xi^{0}_{b} \to  \pi^{0} \Lambda^{0}$ }&${-1 \over \sqrt{3} }$ (0  & -1 & -1 & 5 & 5 & -1 & -1 &3 & -1 & -3 & 5)\\
& ${1 \over 2\sqrt{3} }$ ( 2& 1& 1&-1 &-1 &2 &-1 &-1 &1 &1 & \\
&-4 &-4 &0 &-2 &-2 &-8 &-13 &3 &-3 &13) & \\
\hline
\multirow{3}{*}
{$\Xi^{0}_{b} \to  \pi^{0} \Sigma^{0}$} & -(2  & -1 & -1 & 1 & 1 & 1 & 1 &1 & 1 & -5 & -5)\\
&${1 \over 2 }$ ( 0&-1 &-1 &1 &1 &2 &-1 & -1& 1& 1& \\
&0 &0 &0 & -2& -2&0 &-1 &-1 &1 &1 )& \\
\hline
\hline
\end{tabular}
\label{tab:B2S0}
\end{center}
\end{table}
\begin{table}[h!]
\begin{center}
\caption{$SU(3)$ decay amplitudes for $\Delta S=-1, \Xi^{0}_{b} \to MP$ processes.}
\begin{tabular}{l|ccccccccccc}
\hline\hline
\multirow{3}{*}
 {Decay mode} & $a(\overline{3})$ & $a(6)_{1}$ &  $a(6)_{2}$ & $a(\overline{15})_{1}$ & $a(\overline{15})_{2}$ & $b(\overline{3})_{1}$
 & $b(\overline{3})_{2}$  & $b(6)_{1}$ & $b(6)_{2}$ & $b(\overline{15})_{1}$ & $b(\overline{15})_{2}$\\&$c(\overline{3})$ & $d(\overline{3})_{1}$ & $d(\overline{3})_{2}$ & $e(\overline{3})_{1}$ & $e(\overline{3})_{2}$ &
 $c(6)$ & $d(6)_{1}$ & $d(6)_{2}$ & $e(6)_{1}$ & $e(6)_{2}$& \\ &
 $f(6)$ & $g(6)$ & $m(6)$ & $n(6)_{1}$ & $n(6)_{2}$ & $c(\overline{15})$ & $d(\overline{15})_{1}$ & $d(\overline{15})_{2}$ &$e(\overline{15})_{1}$ & $e(\overline{15})_{2}$&\\
\hline
\multirow{3}{*}
{$\Xi^{0}_{b} \to \pi^{+} \Xi^{-}$ }& -2(0 & 0 & -1 & 0 & -1  & 1 & 0 & 1 & 0 & 3 & 0)\\
&(-1 &0 &-1 &1 &0 &1 &-1 &0 &0 &1 & \\
&-2 &-2 &0 &-2 &0 &3 &3 &-2 &2 &-3) & \\
\hline
\multirow{3}{*}
{$\Xi^{0}_{b} \to K^{-} \Sigma^{+} $} & -2(0 & -1& 0 & -1 & 0 & 0 & 1 & 0 & 1 & 0 & 3)\\
&(-1 &-1 &0 &0 &1 &1 &0 &-1 &1 &0 & \\
&2 &2 &0 &0 &-2 &-3 &-2 &3 &-3 &2) & \\
\hline
\multirow{3}{*}
{$\Xi^{0}_{b} \to \eta_{8} \Xi^{0} $} &  ${-2 \over \sqrt{6}}$ (0 & 2& -1 & 2 & -1 & 1 & -2 & -1 & 0 & -1 & 6)\\
&${1 \over \sqrt{6}}$(1 &2 &-1 &1 &-2 &-1 &-1 &2 &-2 &1 & \\
&-2 &-2 &0 &-2 &4 &-7 &-9 &8 &-8 &9) & \\
\hline
\multirow{3}{*}
{$\Xi^{0}_{b} \to \bar{K}^{0} \Lambda^{0} $} &  ${-2 \over \sqrt{6}}$ (0 & -1& 2 & -1 & 2 & -2 & 1 & 0 & -1 & 6 & -1)\\
&${1 \over \sqrt{6}}$(1 &-1 &2 &-2 &1 &-1 &2 &-1 &1 &-2 & \\
& 2&2 &0 & 4& -2& 7&8 &-9 &9 &-8) & \\
\hline
\multirow{3}{*}
{$\Xi^{0}_{b} \to \bar{K}^{0} \Sigma^{0} $} &  ${-2 \over \sqrt{2}}$ (0 & 1& 0 & 1 & 0 & 0 & -1 & -2 & 1 & 4 & 1)\\
&${1 \over \sqrt{2}}$(1 &1 & 0&0 &-1 &-1 &0 &1 &-1 &0 & \\
& 2&2 &0 &0 &2 &3 &2 &-3 &3 &-2) & \\
\hline
\multirow{3}{*}
{$\Xi^{0}_{b} \to \pi^{0} \Xi^{0} $} &  ${-2 \over \sqrt{2}}$ (0 & 0& 1 & 0 & 1 & -1 & 0 & 1 & -2 & 1 & 4)\\
& ${1 \over \sqrt{2}}$ (1& 0& 1& -1& 0& -1& 1&0 &0 &-1 & \\
&-2 &-2 &0 &2 &0 &-3 &-3 &2 &-2 &3) & \\
\hline
\hline
\end{tabular}
\label{tab:B2S1}
\end{center}
\end{table}

\begin{table}[h!]
\begin{center}
\caption{$SU(3)$ decay amplitudes for $\Delta S=0, \Lambda^{0}_{b} \to MP$ processes.}
\begin{tabular}{l|ccccccccccc}
\hline\hline
\multirow{3}{*}
 {Decay mode} & $a(\overline{3})$ & $a(6)_{1}$ &  $a(6)_{2}$ & $a(\overline{15})_{1}$ & $a(\overline{15})_{2}$ & $b(\overline{3})_{1}$
 & $b(\overline{3})_{2}$  & $b(6)_{1}$ & $b(6)_{2}$ & $b(\overline{15})_{1}$ & $b(\overline{15})_{2}$\\&$c(\overline{3})$ & $d(\overline{3})_{1}$ & $d(\overline{3})_{2}$ & $e(\overline{3})_{1}$ & $e(\overline{3})_{2}$ &
 $c(6)$ & $d(6)_{1}$ & $d(6)_{2}$ & $e(6)_{1}$ & $e(6)_{2}$& \\ &
 $f(6)$ & $g(6)$ & $m(6)$ & $n(6)_{1}$ & $n(6)_{2}$ & $c(\overline{15})$ & $d(\overline{15})_{1}$ & $d(\overline{15})_{2}$ &$e(\overline{15})_{1}$ & $e(\overline{15})_{2}$&\\
\hline
\multirow{3}{*}
{$\Lambda^{0}_{b} \to K^{+} \Sigma^{-}$} & 2(0 & 0 & -1 & 0 & -1  & 1 & 0 & 1 & 0 & 3 & 0)\\
&(1 &0 &1 &-1 &0 &1 &1 &0 &0 &-1 &\\
&2 &2 &0 &2 &0 &-3 &-3 &2 &-2 & 3)&\\
\hline
\multirow{3}{*}
{$\Lambda^{0}_{b} \to \pi^{-} p$} &2 (0 & -1 & 0 & -1 & 0  & 0 & 1 & 0 & 1 & 0 & 3)\\
&( 1& 1& 0& 0&-1 & 1&0 &1 &-1 &0 &\\
&-2 &-2 &0 &0 &2 &3 &2 &-3 &3 &-2 &\\
\hline
\multirow{3}{*}
{$\Lambda^{0}_{b} \to \eta_{8} n $} &  $ {2 \over \sqrt{6}}$(0 & -1 & 2 & -1 & 2  & -2 & 1 & 2 & -3 & 2 & 3)\\
& ${1 \over \sqrt{6}}$( -1& 1& -2& 2& -1& 1&-2 &1 &-1 &2 &\\
&2 &2 &0 &-4 &2 & 1& 0& 1&-1 & 0&\\
\hline
\multirow{3}{*}
{$\Lambda^{0}_{b} \to K^{0} \Lambda^{0} $} &  $ {2 \over \sqrt{6}}$(0 & 2 & -1 & 2 & -1  & 1 & -2 & -3 & 2 & 3 & 2)\\
& ${1 \over \sqrt{6}}$( -1 &-2 & 1& -1& 2&1 &1 &-2 &2 &-1 &\\
& -2& -2& 0& 2& -4&-1 &1 &0 &0 &-1) &\\
\hline
\multirow{3}{*}
{$\Lambda^{0}_{b} \to K^{0} \Sigma^{0} $} &  $ {2 \over \sqrt{2}}$(0 & 0 & 1 & 0 & 1  & -1 & 0 & -1 & 0 & 5 & 0)\\
&$ {1 \over \sqrt{2}}$(-1 &0 &-1 & 1& 0&1 &-1 &0 &0 &1 &\\
&-2 &-2 &0 &-2 &0 &-5 &-5 &6 &-6 & 5)&\\
\hline
\multirow{3}{*}
{$\Lambda^{0}_{b} \to \pi^{0} n $ }&  $ {2 \over \sqrt{2}}$(0 & 1& 0 & 1 & 0  & 0 & -1 & 0 & -1 & 0 & 5)\\
&$ {1 \over \sqrt{2}}$( -1& -1& 0&0 &1 &1 &0 &-1 &1 &0 &\\
& 2&2 &0 &0 &-2 &5 &6 &-5 & 5&-6) &\\
\hline
\hline
\end{tabular}
\label{tab:B3S0}
\end{center}
\end{table}

\begin{table}[h!]
\begin{center}
\caption{$SU(3)$ decay amplitudes for $\Delta S=-1, \Lambda^{0}_{b} \to MP$ processes.}
\begin{tabular}{l|ccccccccccc}
\hline\hline
\multirow{3}{*}
 {Decay mode} & $a(\overline{3})$ & $a(6)_{1}$ &  $a(6)_{2}$ & $a(\overline{15})_{1}$ & $a(\overline{15})_{2}$ & $b(\overline{3})_{1}$
 & $b(\overline{3})_{2}$  & $b(6)_{1}$ & $b(6)_{2}$ & $b(\overline{15})_{1}$ & $b(\overline{15})_{2}$\\&$c(\overline{3})$ & $d(\overline{3})_{1}$ & $d(\overline{3})_{2}$ & $e(\overline{3})_{1}$ & $e(\overline{3})_{2}$ &
 $c(6)$ & $d(6)_{1}$ & $d(6)_{2}$ & $e(6)_{1}$ & $e(6)_{2}$& \\ &
 $f(6)$ & $g(6)$ & $m(6)$ & $n(6)_{1}$ & $n(6)_{2}$ & $c(\overline{15})$ & $d(\overline{15})_{1}$ & $d(\overline{15})_{2}$ &$e(\overline{15})_{1}$ & $e(\overline{15})_{2}$&\\
\hline
\multirow{3}{*}
{$\Lambda^{0}_{b} \to K^{+} \Xi^{-}$} & 2(1 & -1 & 0 & 3 & -2  & 1 & 0 & 1 & 0 & 3 & 0)\\
&(0 &0 &1 &-1 &0 &-1 &0 &1 &-1 &0 &\\
&0 &0 &0 &0 &2 &-3 &0 &1 &-1 &0 )&\\
\hline
\multirow{3}{*}
{$\Lambda^{0}_{b} \to \pi^{+} \Sigma^{-}$} &2 (1 & -1 & 1 & 3 & -1  & 0 & 0 & 0 & 0 & 0 & 0)\\
&(-1 &0 &0 &0 &0 &0 &-1 &1 &-1 & 1&\\
&-2 &-2 &0 &-2 &2 &0 &3 &-1 &1 &-3) &\\
\hline
\multirow{3}{*}
{$\Lambda^{0}_{b} \to K^{-} p$ }&2 (1 & 0 & -1 & -2 & 3  & 0 & 1 & 0 & 1 & 0 & 3)\\
&(0 &1 &0 &0 &-1 &-1 &1 &0 &0 & -1&\\
& 0&0 &0 &2 &0 &3 &1 &0 &0 &-1 )&\\
\hline
\multirow{3}{*}
{$\Lambda^{0}_{b} \to \pi^{-} \Sigma^{+}$} & 2(1 & 1 & -1 & -1 & 3  & 0 & 0 & 0 & 0 & 0 & 0)\\
&(-1 &0 &0 &0 &0 &0 &1 &-1 &1 &-1 &\\
& 2&2 &0 &2 &-2 &0 &-1 &3 &-3 & 1)&\\
\hline
\multirow{3}{*}
{$\Lambda^{0}_{b} \to K^{0} \Xi^{0}$ }& 2(1 & 1 & 0 & -1 & -2  & 1 & 0 & -1 & 0 & -1 & 0)\\
&( 0& 0& 1&-1 &0 &1 &0 &-1 &1 &0 &\\
& 0&0 &0 &0 &-2 &1 &0 &-3 &3 &0) &\\
\hline
\multirow{3}{*}
{$\Lambda^{0}_{b} \to \bar{K}^{0} n$} & 2(1 & 0 & 1 & -2 & -1  & 0 & 1 & 0 & -1 & 0 & -1)\\
&( 0&1 &0 &0 &-1 &1 &-1 &0 &0 &1 &\\
& 0&0 &0 &-2 &0 &-1 &-3 &0 &0 &3) &\\
\hline
\multirow{3}{*}
{$\Lambda^{0}_{b} \to \pi^{0} \Sigma^{0}$} & 2(1 & 0 & 0 & 1 & 1  & 0 & 0 & 0 & 0 & 0 & 0)\\
& (-1&0 &0 &0 &0 & 0&0 &0 &0 &0 &\\
&0 &0 &0 &0 &0 &0 &1 &1 &-1 &-1) &\\
\hline
\multirow{3}{*}
{$\Lambda^{0}_{b} \to \eta_{8} \Lambda^{0}$} & ${2\over3}$ (3 & 0 & 0 & -3 & -3  &2 & 2 & 0 & 0 & -6 & -6)\\
&${1\over3}$ (1 &2 &2 &-2 &-2 &0 &0 &0 &0 &0 &\\
&0 &0 & 0&0 &0 &0 &-3 &-3 &3 &3) &\\
\hline
\multirow{3}{*}
{$\Lambda^{0}_{b} \to \eta_{8} \Sigma^{0}$} & ${2 \over \sqrt{3}}$ (0 & -1 & -1 & 2 & 2  & 0 & 0 & 2 & 0 & -4 & 0)\\
&${1 \over \sqrt{3}}$ ( 0&0 &0 &0 &0 &-2 &1 &1 &-1 &-1 &\\
&2 &2 &0 &2 &2 &4 &6 &-2 &2 &-6) &\\
\hline
\multirow{3}{*}
{$\Lambda^{0}_{b} \to \pi^{0} \Lambda^{0}$} & ${2 \over \sqrt{3}}$ (0 & -1 & -1 & 2 & 2  & 0 & 0 & 0 & 2 & 0 & -4)\\
&${1 \over \sqrt{3}}$ (0 &0 &0 &0 &0 &-2 &1 &1 &-1 &-1 &\\
& -2& -2& 0&2 &2 &-4 &2 & 6&-6 &2) &\\
\hline
\hline
\end{tabular}
\label{tab:B3S1}
\end{center}
\end{table}

\end{document}